\documentclass[prb,aps,amsfonts,amssymb,nofootinbib,twocolumn,amsmath,superscriptaddress]{revtex4-2}
\usepackage{graphicx}
\usepackage{bm}
\usepackage{hyperref}
\usepackage{float}
\usepackage{siunitx}
\usepackage{xcolor}
\usepackage{natbib}
\usepackage{mhchem}
\begin{document}

\title{Derivation and application of a general scaling relation between the dc and asymptotic high-frequency optical Hall responses}

\author{E. Abelev}
\affiliation{William H. Miller III Department of Department of Physics and Astronomy, The Johns Hopkins University, Baltimore, Maryland 21218, USA}

\author{R. Romero III} 
\affiliation{William H. Miller III Department of Department of Physics and Astronomy, The Johns Hopkins University, Baltimore, Maryland 21218, USA}

\author{N. P. Armitage}
\email{npa@jhu.edu}
\affiliation{William H. Miller III Department of Department of Physics and Astronomy, The Johns Hopkins University, Baltimore, Maryland 21218, USA}

\begin{abstract}

Based on the Kramers-Krong relations, we derive and  apply a quite general expression that relates the low frequency  quasi-dc Hall conductivity to the asymptotic high frequency optical Hall response 
(and its manifestations in Kerr and Faraday effects) for time-reversal symmetry breaking (TRSB) states of matter like ferromagnets and time-reversal symmetry breaking superconductors as well as metals in magnetic field.  Parametric plots shows this relation is obeyed exactly for the trivial single-mode case of sharp cyclotron resonance, and approximately for theoretical models for ferromagnets and TRSB superconductors.  We also apply it to the experimental data from a variety of ferromagnetic systems and show reasonable agreement there as well.  We use the relation to predict, from the size of the spontaneous Kerr effect at the pseudogap temperature of the cuprate superconductors that cuprates should exhibit an anomalous Hall effect of approximately 0.1 Ohm$^{-1}\cdot$cm$^{-1}$.  This is a small value, but one within experimental reach and we encourage the search for it.  Although not explicitly quantum geometric, our treatment has some similarities to efforts to set bounds on physical quantities based on quantum geometric relations and limited physical information.

\end{abstract}

\maketitle

\section{Introduction}

Spontaneous time-reversal symmetry breaking (TRSB) in interacting states of matter is a topic of long term interest.  It can be apparent in macroscopic response optical functions when the product of the time-reversal symmetry operation and a lattice translation is not a symmetry of the system~\cite{dzyaloshinskii1991space,armitage2014constraints}.  For conductors, TRSB manifests as a Hall conductivity, which is symmetry allowed at all frequencies.  The dc anomalous Hall effect is its most obvious manifestation~\cite{nagaosa2010anomalous} and the magneto-optical Kerr effect (MOKE) is another.  It is related to the magnitude of the Hall response in the optical range~\cite{Xia2006,kapitulnik2009polar}.  Although Kerr and Faraday effect experiments have been done for many years~\cite{Kato,Keszmarki,Kim,Krinchuk}, new techniques have been developed that allow measurements of the finite frequency Hall conductivity in regimes of sensitivity and frequency that were not possible previously.   Zero-loop Sagnac experiments allow measurements at 1550 nm with 10 $\mu$W of power with sensitivity as high as $1\times10^{-7}\text{ rad/}\sqrt{\text{Hz}}$~\cite{Xia2006,kapitulnik2009polar}.  Recent developments in the much lower THz frequency range now allow extremely precise measurements of the THz Faraday rotation, which is also a measure of the Hall response~\cite{tagayHighprecisionMeasurementsTerahertz2024,romeroTerahertzRangePolarization2025}.   Such experiments have been used to reveal important aspects of ferromagnets, commensurate antiferromagnets, and TRSB superconductors~\cite{morris2012polarization, bhandiaTHzrangeFaradayRotation2020, chauhanMeasurementsCyclotronResonance2022, lubashevskyOpticalBirefringenceDichroism2014, mukherjeeLinearDichroismInfrared2020, lauritaLowenergyMagnonDynamics2017}.

In this work we derive and apply a general relation using the Kramers-Kronig relations that connects the low-frequency (quasi-dc) Hall response to the high-frequency optical Hall response in time-reversal symmetry–breaking (TRSB) states of matter, including ferromagnets, TRSB superconductors, and conductors in an applied magnetic field.  We derive this expression for a particular form of a TRSB superconductor, but the expression applies equally to magnetic metals or conductors in magnetic field.  It is exact in the limit where the spectrum for the dissipative part of $\sigma_{xy}$ is given by a single mode, but as we show can be applied more generally to reasonable approximation even in more realistic models.   We then apply the relation to experimental data from a variety of ferromagnetic systems and where data is available and find reasonable agreement in parametric plots that compare the high frequency Hall conductivity and the dc Hall effect. Finally, we use this relation to predict that the spontaneous Kerr effect observed in cuprate superconductors should be accompanied by an anomalous Hall effect, with a magnitude that appears to be within experimentally accessible limits.  Although not explicitly quantum geometric in origin, our treatment has some connection to efforts extracting physical quantities based on quantum geometric relations and limited physical information~\cite{souza2025optical,komissarov2024quantum,onishi2025quantum,verma2025instantaneous,onishi2024universal}.

\section{Background}

We begin with a general Kramers-Kronig (KK) expression for the real part of the Hall conductivity, $\sigma_{1, xy}$ i.e.,
\begin{equation}
    \sigma_{1, xy}(\omega) = -\frac{2}{\pi} \int_0^\infty \frac{\omega' \sigma_{2,xy}(\omega')}{\omega'^2 - \omega^2}d\omega'.
    \label{eq1}
\end{equation}

The Kramers-Kronig relations are predicated on the assumption of linearity, causality, and analyticity in the upper half of the complex frequency plane.  One can find the non-dissipative (dissipative) conductivity at any frequency if we know the dissipative (non-dissipative) conductivity for all frequencies.  The dissipative part of the Hall conductivity can be expressed as the difference between dissipative conductivities for right and left circularly polarized light ($\sigma_{r,l})$.  The imaginary part of the Hall conductivity $\sigma_{2,xy}$ is the dissipative response of the Hall response and can be expressed as
\begin{equation}
    \sigma_{2, xy} = \Im\left(\frac{\sigma_r-\sigma_l}{2i}\right) = -\left(\frac{\sigma_{1r}-\sigma_{1l}}{2}\right) .
    \label{eq2}
\end{equation}

Substituting  (\ref{eq2}) $\rightarrow$ (\ref{eq1}) one has, 

\begin{equation*}
    \sigma_{1, xy} (\omega)  = \frac{1}{\pi}\int_0^\infty \frac{\omega'\left[\sigma_{1r}(\omega')-\sigma_{1l}(\omega')\right]}{\omega'^2 - \omega^2}d\omega'.
\end{equation*}

\begin{figure}
    \centering \; 
    \includegraphics[width=\linewidth]{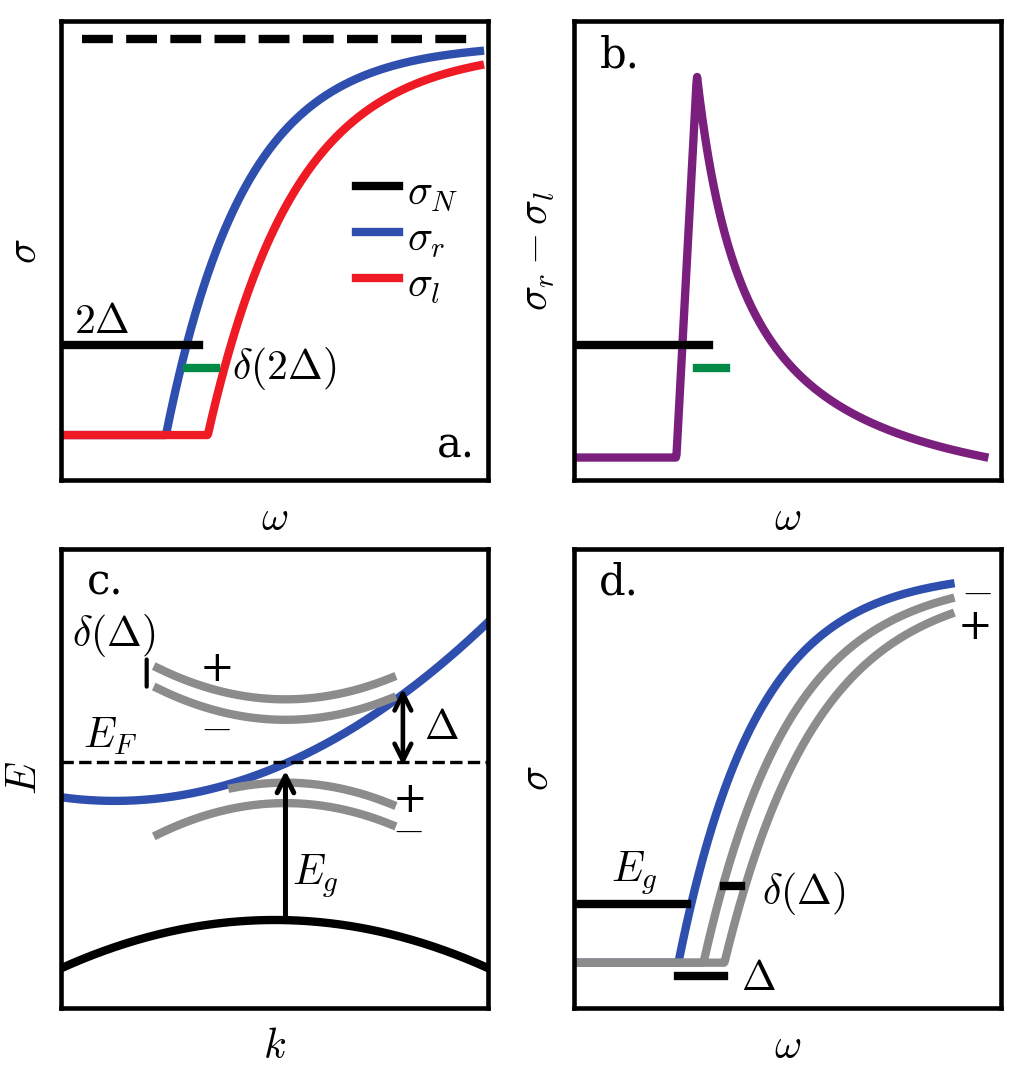}
    \caption{a) In the dirty limit TRSB superconductor case the real part of the conductivity may manifest a different superconducting gap for the right and left absorptive channels. The energy scales, $2\Delta$ and $\delta(2\Delta)$ are denoted in black and green respectively. b) The difference of the conductivities shown in a). When this quantity is finite it is of order the normal state conductivity $\sigma_N$, but falls off quickly with increasing frequency. c) Schematic of the band structure for TRSB SC in a multiband system. The grey bands could be spin-split Bogoliubon branches which modify the optical conductivity as shown in. d) The optical conductivity for this case.  }
    \label{Fig1}
\end{figure}

As this is an absorptive response, we expect that $[\sigma_{1r}-\sigma_{1l}]$ is only non-zero in a limited  region in frequency.   As shown in Figs.~\ref{Fig1}a) and b) this would be case, for instance, in a TRSB superconductor in the dirty limit where there could be different superconducting gaps apparent in $\sigma_r$ and $\sigma_l$, the difference of which we take as $\delta (2\Delta) = 2\Delta_r- 2\Delta_l$ with an average value of $2\Delta$.  (Here we are using units where $\Delta$ is expressed in angular frequency i.e. energy divided by $\hbar$.)  As shown in Fig.~\ref{Fig1} we expect that this quantity would be non-zero only in a compartively small energy range  $\delta (2\Delta) $ centered around $2 \Delta$~\cite{Brydon,Can,Gradhand,Huang,Taylor,Lutchyn}. This is the key assumption of this analysis, and errors (which we argue to be small in most realistic cases) are introduced when this is not the case.  As we discuss below the same analysis essentially applies for a clean multi-gap TRSB superconductor or ferromagnets where interband transitions are gapped or shifted by the occurrence of TRSB state.

Note that in in the frequency region where the quantity $  \sigma_{2, xy} $ is non-zero, we expect it to be of order $\sigma_N$ the conductivity in the normal state.  We perform the KK transform to find the high frequency limit, which is defined as frequencies well above the region where  $  \sigma_{2, xy} $  is finite.  As noted above, we expect that $\sigma_{1  r} - \sigma_{1 l} $ is appreciably different from zero only in a narrow frequency range $\delta (2 \Delta)$ for  $ \omega' \approx 2\Delta$.    Therefore for $\omega \gg 2\Delta$, we can write

\begin{equation}
     \sigma_{1, xy}(\omega \gg 2\Delta )  \sim  - \frac{1}{\pi}\frac{1}{\omega^2}\int_0^\infty  \omega' [\sigma_{1  r} - \sigma_{1 l} ] d\omega'.
    \label{eq3}
\end{equation}

 Furthermore, because we are considering the case $\delta \Delta \ll \Delta$, the $\omega'$ inside the integral can be taken to be its mean value in the region where  $  \sigma_{2, xy} \neq 0 $ and pulled outside the integral, i.e.
 
 \begin{align}
     \sigma_{1, xy}(\omega \gg 2\Delta ) \sim  - \frac{1}{\pi}\frac{2\Delta }{\omega^2} \int_0^\infty [ \sigma_{1  r} - \sigma_{1 l} (\omega')] d\omega'.
\end{align}
 
The integral is now simply the differences of spectral weights between the right and left absorptive channels.  Putting it all together we have

 \begin{align}
      \sigma_{1, xy}(\omega \gg 2\Delta )  \sim - \frac{1}{\pi}\frac{2\Delta }{\omega^2}  \gamma \sigma_N \delta(2  \Delta) .
      \label{HighFreq}
\end{align}

Here $\gamma$ is a factor that accounts for the modifications of the optical matrix elements for above the gap absorptions.  For the case of a dirty limit BCS superconductor $\gamma = \pi/2$.  To recap, Eq.~\ref{HighFreq} is an expression that is valid for frequencies well above the region of absorption ($\omega \gg 2\Delta$) if the off-diagonal absorption is found in a narrow range of frequencies (e.g. $\delta(2 \Delta ) \ll2 \Delta$).

We can derive an analogous expression for the low frequency limit, i.e.  $\omega \ll 2\Delta$,
\begin{align*}
    \sigma_{1, xy}(\omega \rightarrow 0) & =   \frac{1}{\pi}\int_0^\infty\frac{\omega' [ \sigma_{1  r} - \sigma_{1 l}] }{\omega'^2(1-\frac{\omega^2}{\omega'^2})}d\omega',\\
    & \sim  \frac{1}{\pi}\int_0^\infty\frac{\sigma_{1  r} - \sigma_{1 l}}{\omega'}[1+\frac{\omega^2}{\omega'^2}]d\omega',\\
    &\sim  \frac{1}{\pi} \frac{1}{2\Delta}  (1+\frac{\omega^2}{ (2\Delta )^2})\int_0^\infty  [\sigma_{1  r} - \sigma_{1 l} ] d\omega'.\\
\end{align*}

Here we have again used the fact that the integral is only non-zero in a narrow frequency around $2\Delta$ and that $\omega'$ can be replaced in the integral by its mean value and then taken out as a constant.  And the integral is again the difference in absorption for right and left circularly polarized light.

\begin{align}
   \sigma_{1, xy} (\omega \rightarrow 0) &\sim  \frac{1}{\pi} \frac{  \gamma \sigma_N \delta(2  \Delta)}{2\Delta}  (1+\frac{\omega^2}{ (2\Delta )^2}).
\end{align}

Note this form is constant as  $\omega\rightarrow0$ with a  quadratic correction that is small as long as the frequencies being considered are well below the gap scale.  It is

\begin{equation}
    \sigma_{1, xy}(0)=  \frac{1}{\pi}\frac{\gamma\sigma_N \delta(2\Delta)}{2\Delta}.
    \label{LowFreq}
\end{equation}

We can combine Eqs.~\ref{HighFreq} and ~\ref{LowFreq} to get our key result, that is

\begin{align}
\sigma_{1, xy}(\omega \gg 2\Delta )= - \sigma_{1, xy}(0) \left(\frac{2\Delta}{ \omega}\right)^2.
    \label{DirtyLimit}
\end{align}

This is a general and useful result, which can be used to relate the low frequency Hall response to the high frequency response.  It is valid for frequencies well away from the range of optical absorptions that are different for right and left hand light and depends only on the validity of the KK relations.   It is most accurate for dissipative Hall conductivities that are distributed narrowly in frequency and symmetric, but we find the errors to be only of order unity when these conditions are not met.  Note the minus sign in Eq.~\ref{DirtyLimit}, as the low frequency Hall effect has the opposite sign as the high frequency asymptote for case when $\sigma_{1, xy}$ is narrowly concentrated in frequency.  This will be important later on when we analyze experimental data, and as a minimal consideration for being in the high frequency limit, we will only consider data where the high frequency Kerr rotation gives an opposite sign from the dc (or low frequency) Hall effect.

As shown in Figs.~\ref{Fig1}c) and d), for clean multi-band superconductors or ferromagnets, this expression is still valid, with the gap scale $2\Delta$ being replaced by $E_g$ which is the energy of interband transitions that are gapped or shifted by the occurrence of the TRSB ordered state.  Although these cases are microscopically very different from the scenario in Figs.~\ref{Fig1}a) and b), the key assumption from earlier still holds i.e. $\sigma_{1  r} - \sigma_{1 l} $ is nonzero in a small frequency range. 

\section{Theory Examples }

\begin{figure}
    \centering
    \includegraphics[width=\linewidth]{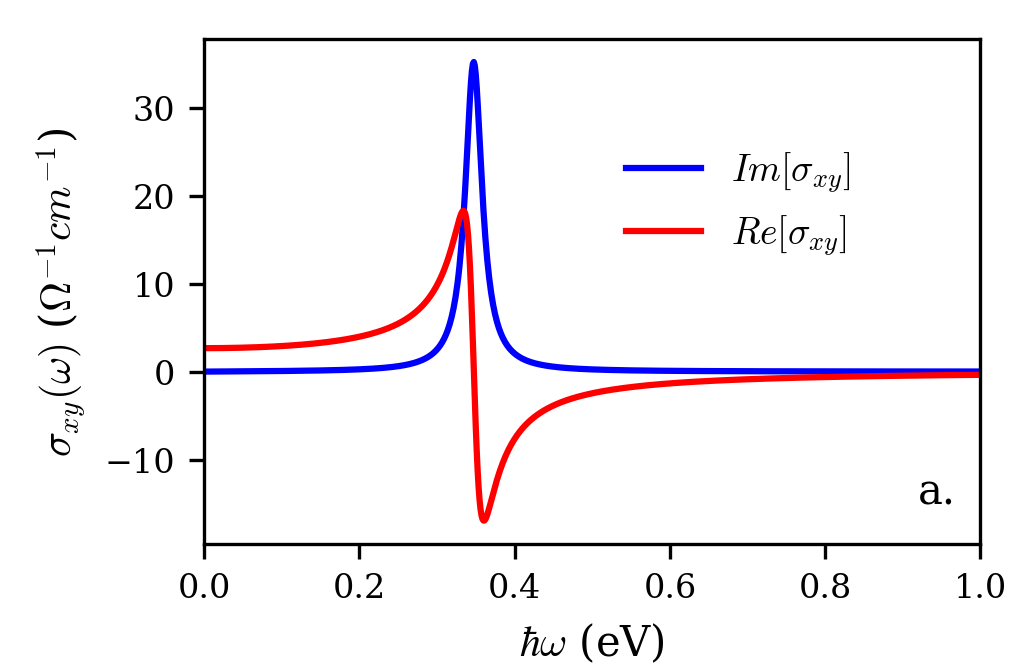}
    \includegraphics[width=\linewidth]{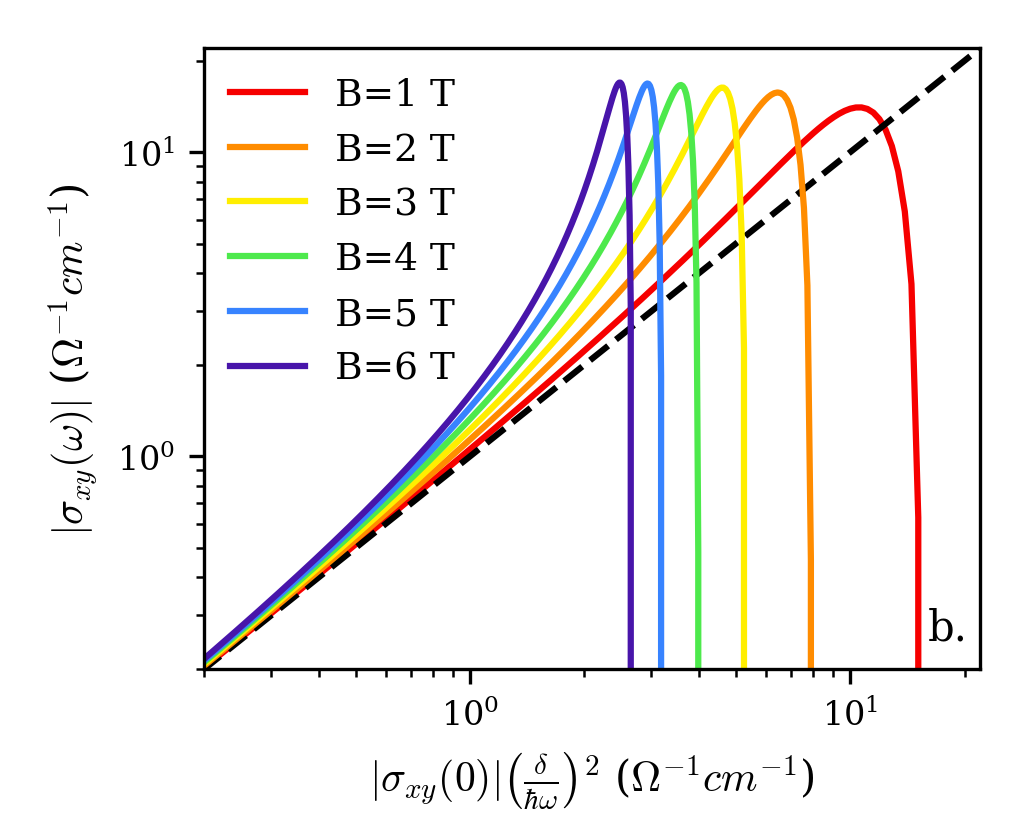}
    \caption{Hall conductivity of cyclotron resonance in Drude model. a) Real and imaginary parts of Hall conductivity spectrum at 6 T. b) Convergence of real part of Hall conductivity from the Drude model with Eq.~\ref{cyclotron2}. $\gamma=2\times10^{13}~s^{-1},~m=2\times10^{-3}~m_e,~n=10^{20}\text{ electrons}/$cm$^3$.   The data approaches the unity line at large $\omega$ (left side of the plot) in accord with Eq.~\ref{cyclotron2}.}
    \label{drude}
\end{figure}

The most straightforward case to which this model should apply is the explicitly single mode situation of a narrow cyclotron resonance.  For a Fermi gas in magnetic field, the zero frequency Drude is shifted to a finite frequency resonance at $\omega_c$.  The real and imaginary parts of the Hall conductivity are 

\begin{align}
   \sigma_{1, xy} (\omega) =  -\frac{ne^2}{m}\frac{\omega_c(\gamma^2-\omega^2+\omega_c^2)}{(\gamma^2-\omega^2+\omega_c^2)^2+4\gamma^2\omega^2} 
      \label{cyclotronReal}
\end{align}

\begin{align}   
   \sigma_{2, xy} (\omega) =  -\frac{ne^2}{m}\frac{2\gamma\omega_c\omega}{(\gamma^2-\omega^2+\omega_c^2)^2+4\gamma^2\omega^2}
   \label{cyclotronImag}
\end{align}
where $\omega_c=\frac{eB}{m}$.  In Fig.~\ref{drude}a) we plot $\sigma_{1, xy}$ and $\sigma_{2, xy}$ as a function of frequency for a cyclotron frequency $\omega_c= 4.37$ eV and damping rate $\gamma = 0.021$ eV. If we expand Eq.~\ref{cyclotronReal} at large $\omega$ we find the expression

\begin{align}
\sigma_{1, xy}(\omega \gg \omega_c,\gamma)= - \sigma_{1, xy}(0) \left(\frac{\omega_c}{\omega}\right)^2.
    \label{cyclotron2}
\end{align}

 Eq.~\ref{cyclotron2} is clearly a version of Eq.~\ref{DirtyLimit} applied to cyclotron motion.  It will be valid at high enough frequencies, but it is an unclear to what extent it can still be a guide even in frequency regions where the dissipation is finite.  In Fig.~\ref{drude}b) we plot {\it parametrically} Eq.~\ref{cyclotronReal} vs. the right side of Eq.~\ref{cyclotron2} on vertical and horizontal axes respectively for a number of values of $\omega_c$.  Here and below we label the energy scale that features in Eq.~\ref{DirtyLimit} and its relatives as $\delta$.  Near the resonance frequency, the data deviates from the high frequency prediction, but it all cases the data approaches the unity line for $\omega \gg \omega_c $ (e.g. the left side of the plot).  The unity line is a graphical expression of Eq.~\ref{cyclotron2} and that the data approaches it for $\omega \gg \omega_c$ indicates  the prediction is valid at large enough  $\omega$.  This shows, perhaps unsurprisingly for such a simple model, that knowing the dc Hall conductivity and the cyclotron resonance frequency is enough to predict the high frequency response.

An important simple model for ferromagnets where the expression does {\it not} work outright is that of a  Dirac continuum model in 2D with TRSB breaking.  For a model of Dirac fermions~\cite{iguchi2009optical,Shimano} with a TRSB breaking mass $m$, scattering rate $\gamma$, and chemical potential $\mu$, the conductivity is 

\begin{equation}
    \sigma_{xy}(\omega,\mu)=\frac{e^2}{8\pi h a}\cdot\frac{m}{\omega+i\gamma}\cdot\ln\frac{-\omega-i\gamma+2\mu}{\omega+i\gamma+2\mu}.
    \label{diraceqn}
\end{equation}

\begin{figure}
    \centering
    \includegraphics[width=\linewidth]{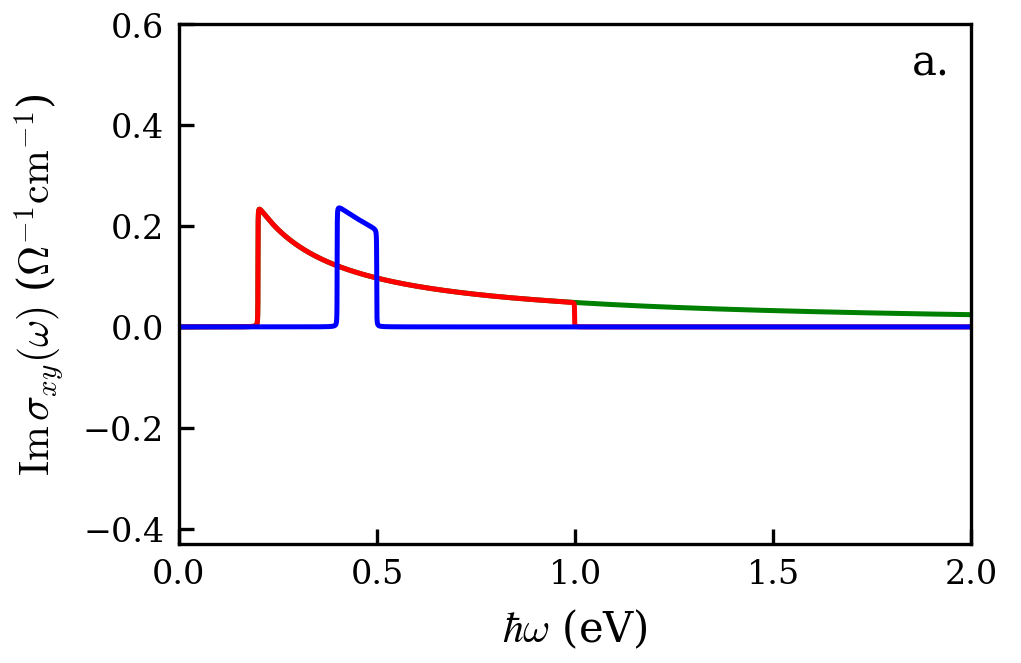}
    \includegraphics[width=\linewidth]{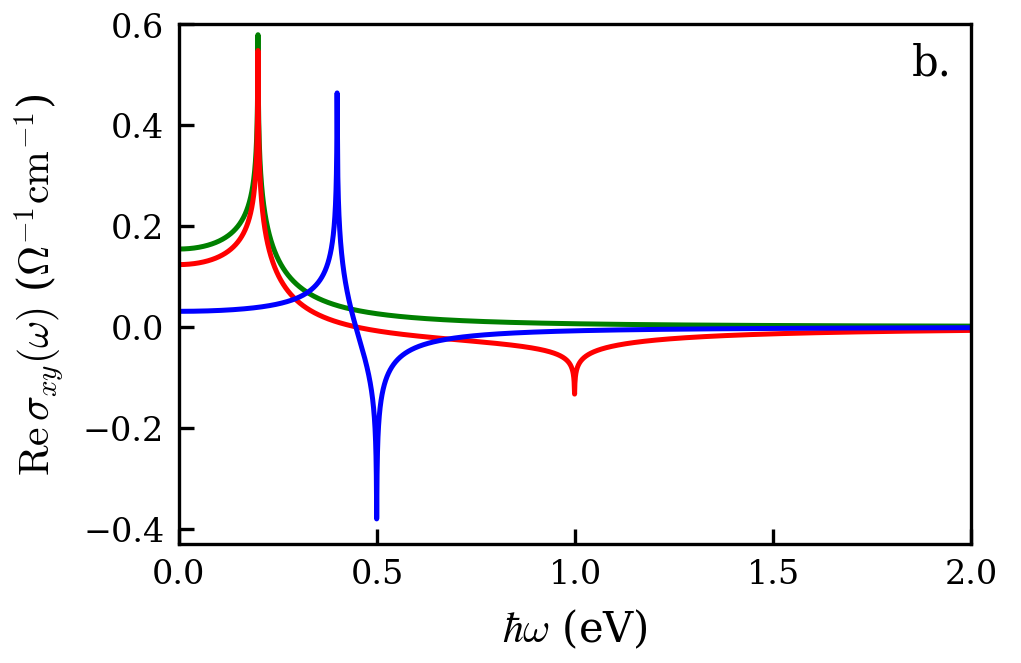}
    \includegraphics[width=\linewidth]{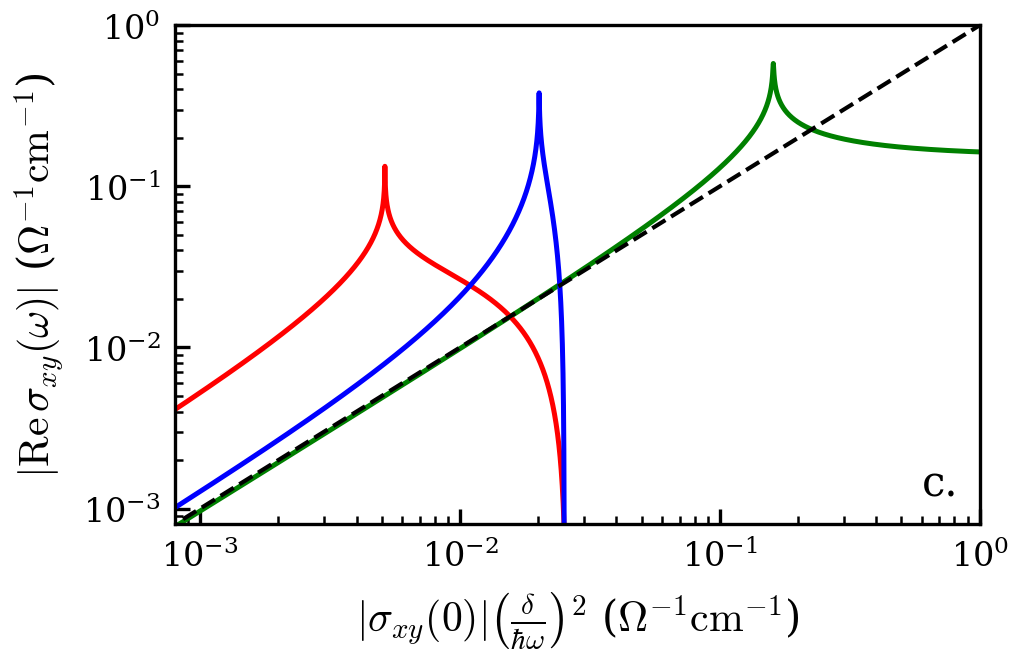}
    \caption{Hall conductivity of truncated and untruncated Dirac cone models. a) Real, b) imaginary parts of conductivity. c) Convergence of real part of Hall conductivity from the Dirac model and Eq.~\ref{DirtyLimit}. Green: untruncated, $m=0.1~eV,~\mu=0.1~eV$. Red: truncated, $m=0.1~eV,~\mu_1=0.1~eV,~\mu_2=0.5~eV$. Blue: truncated, $m=0.2~eV,~\mu_1=0.2~eV,~\mu_2=0.25~eV$}
    \label{diracplot}
\end{figure}

We plot in green in Fig.~\ref{diracplot}a) Eq.~\ref{diraceqn} as a function of frequency.  This continuum model shows a gap from Pauli blocking at 2$\mu$, but also an unphysical lack of high frequency cutoff that allows excitations to arbitrarily large frequency.  Hence, the dissipative response shown in Fig.~\ref{diracplot}a) deviates strongly from a simple pole structure and has a $1/\omega$ power law tail, which continues unabated so Eq.~\ref{DirtyLimit} cannot be valid for $\omega \gg 2\mu $.  Because of the deviation from the simple pole structure and the long tail, as can be seen in Fig.~\ref{diracplot}b), the imaginary part of the Hall response does not change sign as expected from Eq.~\ref{DirtyLimit}.

One way to amend the model to make it more physical is to truncate it, i.e. disregard the contribution from large $k$ to the calculation of $\sigma_{xy}$ from energies past some cutoff $\mu_2$. This corresponds to subtracting off another Hall conductivity of the same form but with a different choice of chemical potential, i.e. 

\begin{equation}
    \sigma_{xy}(\omega)\sim\sigma_{xy}(\omega,\mu_1)-\sigma_{xy}(\omega,\mu_2)
    \label{trunc}
\end{equation}

We plot the imaginary and real parts of the truncated models for various parameters in Figs.~\ref{diracplot}a) and b).  For $\omega \gg \mu_1, \mu_2$, the truncated expression in the parametric plot in Fig.~\ref{diracplot} c) shows a $1/\omega^2$ frequency dependence at large $\omega$ albeit with a prefactor not given precisely by Eq.~\ref{DirtyLimit}.  This is because in Fig.~\ref{diracplot}c) $\delta$ -- in correspondence with our scheme -- was chosen as the peak value, whereas the spectrum is obviously asymmetrically weight towards the high frequency side.  Nevertheless even for the most asymmetric spectrum in red, Eq. ~\ref{DirtyLimit} only underestimates the dependence by a small factor of about 4.   The agreement with the most truncated curve is better as that data is well enough approximated by a single mode.

Interestingly the data from the untruncated model in the parametric plot Fig.~\ref{diracplot}c) also shows agreement with Eq.~\ref{DirtyLimit} if one plots the absolute value of the conductivities, i.e. despite the lack of sign change the high-frequency asymptote goes like $1/\omega^2$ with a coefficient that is precisely proportional to $\sigma_{xy}(0) (2 \mu)^2$ (which one can see by expansion of the real part of Eq.~\ref{diracplot} for high and low $\omega$).   This is a particular anomaly of the Dirac continuum model that we should not expect to be observed in real materials.

As mentioned above, this formalism should apply also to TRSB superconductors as long as one excludes the zero frequency delta-function part of the conductivity.  There is as of yet only one experiment that measures the small, but not zero frequency Hall conductivity~\cite{romeroTerahertzRangePolarization2025}.  Therefore we apply Eq.~\ref{DirtyLimit} to a number of theoretical calculations from the literature for TRSB superconductivity~\cite{Brydon,Can,Huang,Gradhand,Taylor,Lutchyn}.   These encompass a number of different models for different kinds of TRSB superconductors including chiral $d$-wave, chiral $p$-wave, and mixed helical $p$-wave.

We digitized literature plots at discrete nonuniform intervals using WebPlotDigitizer~\cite{WebPlotDigitizer}. Points were selected at higher density near peaks. For experimental data, Hall conductivity, longitudinal conductivity, Kerr rotation and ellipticity, complex refractive index, and complex dielectric function were gathered. Where possible, data was digitized only from exact reported data points, rather than at intervals along a continuous curve. Continuous functions of frequency for optical variables were constructed from data via piecewise interpolation. For Kerr effect data without reported $\sigma_{xy}(\omega)$, the Hall conductivity was derived via the relations

\begin{align}
    \sigma_{xy}(\omega)=-\tilde\theta_K\sigma_{xx}(\omega)&\sqrt{1+i\frac{\sigma_{xx}(\omega)}{\omega\epsilon_0}}\\
    \sigma_{xx}(\omega)= i\epsilon_0 & \omega(1-\tilde n^2).
    \label{kerr}
\end{align}

\begin{figure}
    \centering
    \includegraphics[width=\linewidth]{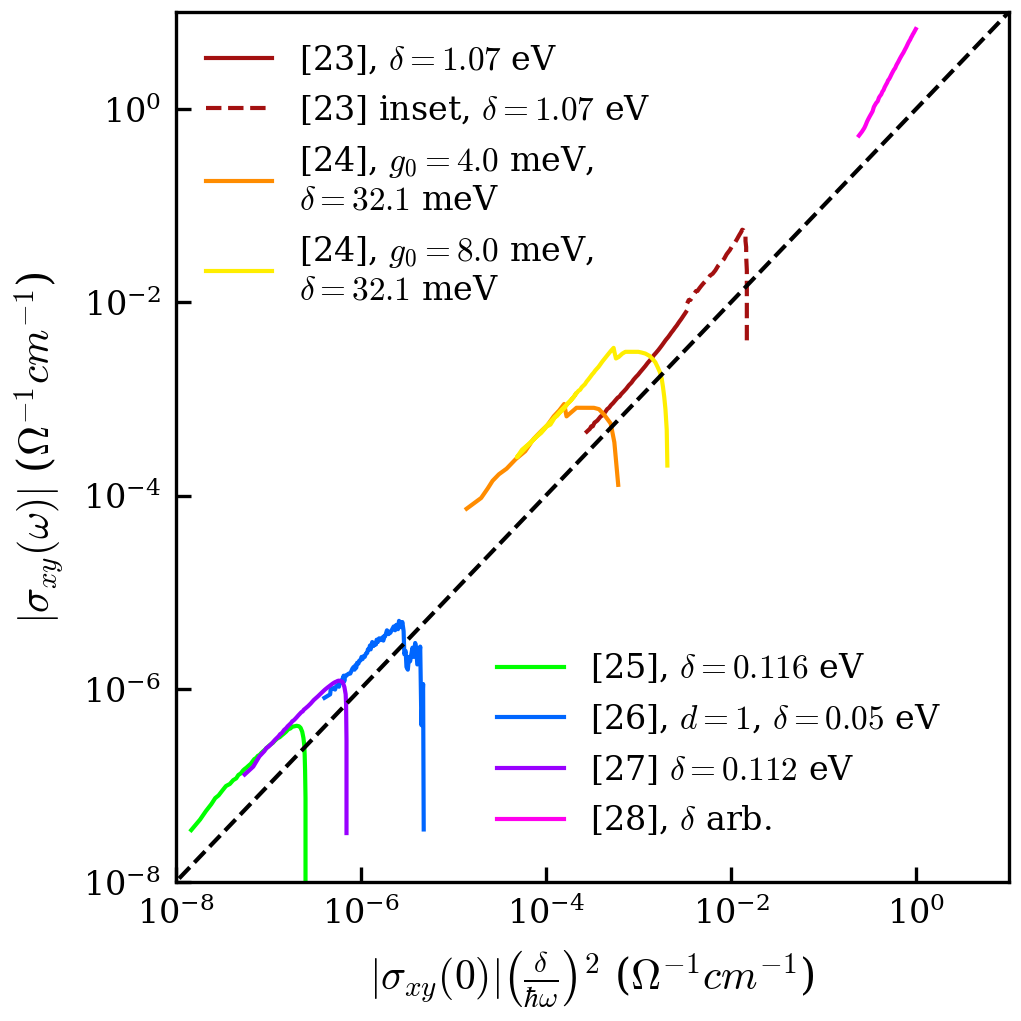}
    \caption{Real part of Hall conductivity predicted theoretically in various superconductors~\cite{Brydon,Can,Gradhand,Huang,Taylor,Lutchyn}, plotted against Eq.~\ref{DirtyLimit}.}
    \label{theory}
\end{figure}

We analyzed only literature data where both real and imaginary parts of the Hall conductivity had been calculated.  Of course the spectra differ greatly from the ideal single mode case.  So the question is --- despite non-idealities --- how good of a guide is Eq.~\ref{DirtyLimit}?   We only included calculations where the the high frequency optical Hall conductivity have an opposite sign than the low frequency Hall effect.  This excluded for instance the calculation of Ref.~\cite{wang2017intrinsic} on UPt$_3$ where a chiral superconducting order parameter in a multi-band system gives both positive and negative dissipative responses and means the low frequency Hall effect is the same sign as the high frequency data plotted.  For the analysis we identify the most prominent peak in the dissipative response with the scale $\delta$.  For data with two prominent resonant features, $\delta$ was selected as the midpoint between the peaks.  Our goal was not precise agreement, but rather investigating if Eq.~\ref{DirtyLimit} provides a rough heuristic.

In Fig.~\ref{theory}, we plot parametrically the theoretical calculations of Re$|\sigma_{xy}(\omega)|$ vs. $|\sigma_{xy}(0)| (\frac{\delta}{ \omega})^2$ in the same fashion as other parametric plots above.  One sees again that there is an asymptotic behavior for larger values of $\omega$.  It is interesting to note that in a manner similar to the red curve of Fig.~\ref{diracplot}, there appears to be a small, but systematic offset, which manifests as a factor between Eq.~\ref{DirtyLimit} and all theory predictions.  This arises from the characteristic asymmetry in dissipative response TRSB superconductors, as seen in Fig. 5 in Ref.~\cite{Lutchyn}, Fig. 2 in Ref.~\cite{Taylor}, and Fig. 2 in Ref.~\cite{Huang}.  For such spectra, spectral weight generally extends in a tail to higher energies than the threshold, which weakly violates the assumptions in the derivation above.  Despite this offset we can conclude that Eq.~\ref{DirtyLimit} provides an good heuristic.

\section{Comparison to Experiment}

We now apply Eq.~\ref{DirtyLimit} to experiments.   There are a handful of experiments on ferromagnetic metals~\cite{Kato,Keszmarki,Kim,Krinchuk}, which have measured the frequency dependent optical Hall conductivity (through Kerr or Faraday rotation) at large enough frequencies to make a comparison.  In real materials there is typically an even broader and sometimes more asymmetric dissipative resonance than expected from theory.  Additionally, weak higher-energy resonances can add effectively constant offsets to lower-energy features and obscure $1/\omega^2$ scaling in the measured spectral range. The layering of opposite-sign resonances at progressively higher resonances could contribute to the ``bumpy" structure of the magnitude of the real spectrum of Hall conductivity that was observed.  Nevertheless in most cases in the literature one finds there to be a broad, but still well-defined feature that can be used to set the energy scale of Eq.~\ref{DirtyLimit}.  Again a necessary requirement for the inclusion of the data set is that the high frequency optical Hall conductivity have an opposite sign than the dc Hall effect.   This is a minimal requirement that the optical data is at high enough frequency to be close to the asymptotic regime.  

 \begin{figure}
    \centering
    \includegraphics[width=\linewidth]{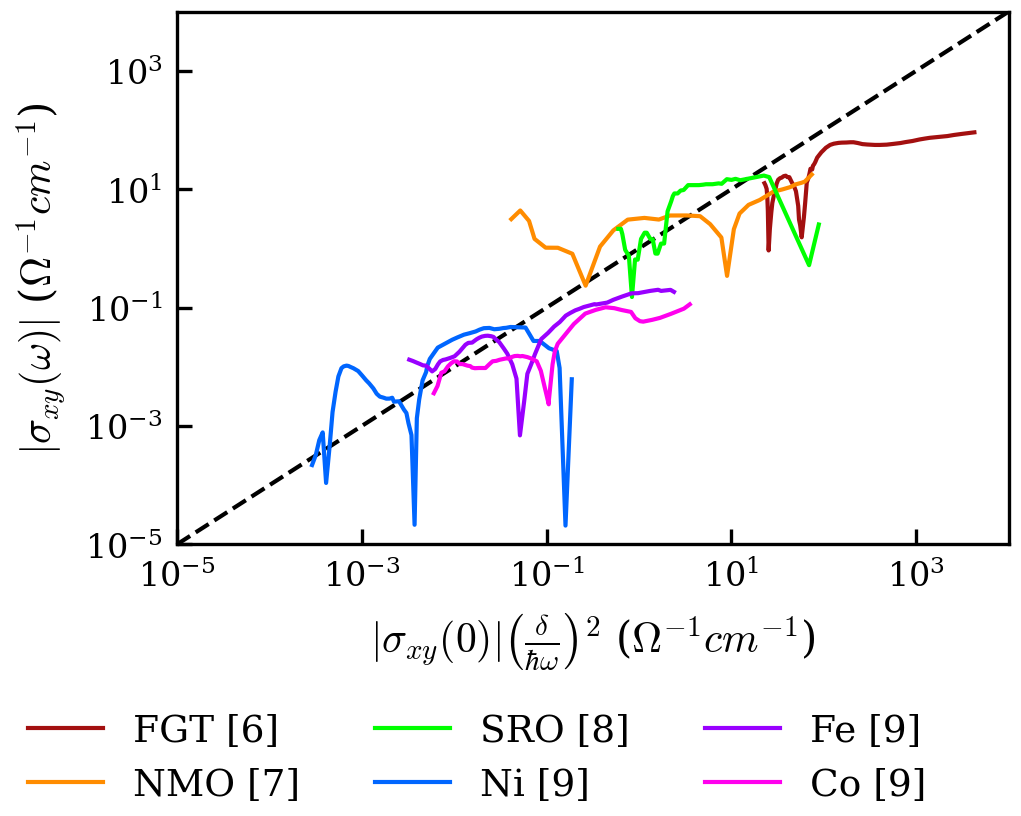}
    \caption{Real part of Hall conductivity experimentally measured in various TRSB ferromagnets~\cite{Kato,Keszmarki,Kim,Krinchuk}.  FGT: \ce{Fe3GeTe2}, NMO: \ce{Nd2Mo2O7}, SRO: \ce{SrRuO3}.}
    \label{experiment}
\end{figure}

We parametrically plot in Fig.~\ref{experiment} experimental results for Re$|\sigma_{xy}(\omega)|$ vs. $|\sigma_{xy}(0)| (\frac{\delta}{ \omega})^2$ for a number of magnetic materials.  Although there is naturally lots of scatter in the data, Eq.~\ref{DirtyLimit} again provides an good guide to the rough scale of the dc Hall effect given the high-frequency optical response or vice versa.  Again we have plotted this data by setting $\delta$ as the energy where Im$|\sigma_{xy}(\omega)|$ reaches a maximum.

\section{The model's predictive power?}

Finally, having confirmed Eq.~\ref{DirtyLimit} for a variety of ideal and non-ideal situations for both theory and experiment we can use this model to make a prediction!   Using a zero-area-loop
Sagnac interferometer and a $\lambda = 1550$ nm (0.8 eV)  laser, Xia \emph{et al.}~\cite{Xia2008} observed a nonzero Kerr
rotation of order $\sim 0.8 \mu$rad in YBa$_2$Cu$_3$O$_{6+x}$ that onsets at a
temperature $T_s$ that tracks the pseudogap temperature $T^{*}$ as a function of doping and extrapolates toward a putative quantum critical point near optimal doping, suggestive of a genuine broken-symmetry transition rather than a crossover~\cite{Kapitulnik2009}.

At least two features of these observations sit awkwardly with thinking of the signal as arising from an ordered state with symmetries consistent with a $c$-axis ferromagnet.  First, the signal was reported to have the same sign measured from opposite faces of the crystal.  Second, it could not be trained by modest fields near $T_c$; its sign appears instead to be fixed by a $\mathcal{T}$-breaking scale already present well above room temperature~\cite{Xia2008,Kapitulnik2009}. These observations motivated a number of $\mathcal{T}$-preserving interpretation in terms of gyrotropic chiral charge order~\cite{Hosur2013}, but those proposals were  withdrawn as it was noted that Onsager reciprocity  forbids a normal-incidence Kerr response in any time-reversal-symmetric medium~\cite{halperin1992hunt,armitage2014constraints}. The conclusion that survives is narrower but firmer --- a finite polar Kerr signal requires broken time-reversal symmetry together with the breaking of all vertical mirrors --- while the microscopic order underlying $T^{*}$ remains unsettled.

Despite these uncertainties (or rather because of them) it makes sense to investigate this physics in different contexts.  If we put aside the possibility that the Hall signal has opposite signs on opposite surfaces, our above discussion indicates that a finite Kerr rotation implies a dc Hall effect.  Going from the scale of the $\theta_K \sim 0.8 \mu$rad YBa$_2$Cu$_3$O$_{6+\delta}$~\cite{Xia2008} and known parameters of the optical response~\cite{basov2005electrodynamics} in the same frequency range one finds a Hall conductivity of approximately 2 $\times 10^{-4}$ Ohm$^{-1}\cdot$cm$^{-1}$ at 0.8 eV.   If we assume a BCS-like value for the energy scale associated from pseudogap onset temperature e.g $\delta = 2 \Delta_{PG}  = 3.5 k T^*$, then we predict that the cuprate superconductors should have a dc anomalous Hall effect of order 0.1 Ohm$^{-1}\cdot$cm$^{-1}$.   This is a small value, but well within the scale of the anomalous Hall effect observed in various systems~\cite{nagaosa2010anomalous}.   The challenge may be that this comes with a very small Hall angle $\theta_H \sim 10^{-5}$ due to the relatively low resistivity of cuprates at low temperature.  However, Hall angles this small have been measured in Germanium~\cite{chazalviel1975spin,sinova2015spin} and in various magnetic metals like SrRuO$_3$ near their sign change temperature~\cite{fang2003anomalous} and so we encourage a search for this effect.  Note that this estimate depends sensitively on the assumed value of $2\Delta_{PG}$ entering Eq.~\ref{DirtyLimit} squared.  A factor of two uncertainty in this scale changes the predicted Hall conductivity by a factor of four. Even allowing for this, the prediction remains within the range of anomalous Hall effects observed in other systems and should be experimentally accessible.

\section{Conclusion}

In this work we have derived a general relation that connects the low-frequency, quasi-dc Hall response to the high-frequency optical Hall response in TRSB states of matter --- ferromagnets, TRSB superconductors, and conductors in a magnetic field. We show that it is obeyed in the most elementary single-mode case of a sharp cyclotron resonance, as well as in theoretical models of ferromagnets and TRSB superconductors.   We also show that it agrees well with experimental data on a variety of ferromagnets. Applying the relation to the Hall conductivity inferred from the Kerr effect observed near the pseudogap temperature of the cuprates, we predict that these systems should likewise exhibit an anomalous Hall effect. The estimated size appears to fall within the measurable range, and we encourage an experimental search for it.

\bigskip

We would like to thank J. Wang for helpful conversations, and G. Reid and J. Xia for careful reading and comments on the manuscript.  This work was supported by the US Department of Energy DE-SC0025245 ``Dynamics and time-evolution in quantum magnets as probed by new nonlinear THz spectroscopies".

\bibliography{main}

\end{document}